\documentclass[twocolumn]{aastex61}
\usepackage{amsmath,amssymb,amstext}
\usepackage{hyperref}

\usepackage{natbib}

\shorttitle{Significance of an excess in a counting experiment}
\shortauthors{G.~Vianello}

\newcommand{\argmax}{\operatornamewithlimits{arg\,max}}
\renewcommand{\textbf}{}
\accepted{\today}

\begin{document}

\title{Significance of an excess in a counting experiment: assessing the impact of systematic uncertainties and the case with Gaussian background}

\correspondingauthor{Giacomo Vianello}
\email{giacomov@stanford.edu}

\author{Giacomo Vianello}
\affiliation{Stanford University, Hansen Experimental Physics Laboratory, \\
450 Serra Mall, \\
Stanford, California 94305}

\begin{abstract}
Several experiments in high-energy physics and astrophysics can be treated as on/off measurements, where an observation potentially containing a new source or effect (``on'' measurement) is contrasted with a background-only observation free of the effect (``off'' measurement). In counting experiments, the significance of the new source or effect can be estimated with a widely-used formula from \citet{LiMa}, which assumes that both measurements are Poisson random variables. In this paper we study three other cases: i) the ideal case where the background measurement has no uncertainty, which can be used to study the maximum sensitivity that an instrument can achieve, ii) the case where the background estimate $b$ in the off measurement has an additional systematic uncertainty, and iii) the case where $b$ is a Gaussian random variable instead of a Poisson random variable. The latter case applies when $b$ comes from a model fitted on archival or ancillary data, or from the interpolation of a function fitted on data surrounding the candidate new source/effect. Practitioners typically use in this case a formula which is only valid when $b$ is large and when its uncertainty is very small, while we derive a general formula that can be applied in all regimes. We also develop simple methods that can be used to assess how much an estimate of significance is sensitive to systematic uncertainties on the efficiency or on the background. Examples of applications include the detection of short Gamma-Ray Bursts and of new X--ray or $\gamma$-ray sources. All the techniques presented in this paper are made available in a Python code ready to use.
\end{abstract}


\section{Introduction}

    The problem of detecting a signal over the background in photon-counting experiments is common in high-energy astronomy and in other fields. \textbf{It has been treated several times in the past} \citep[see for example][and references therein]{Cousins2008}, particularly in the seminal paper by \citet{LiMa}. These authors derived an expression for the significance of a new signal in a on/off measurement: a photon detector observes a region of the sky where a new source is suspected to exist (``on'' region), and then turns to observe a region of the sky which does not contain any source (``off'' region). The background is estimated by counting the events in the off region, and it is assumed that the ``true'' background in the on and in the off region is the same after correcting for the known difference in efficiency. Under these circumstances, the number of counts observed in the on and off regions are both random variables with a Poisson probability density function. 

    In this paper we treat three different situations which are interesting in astronomical and other applications. 
 
The first scenario involves an ideal situation where the background is known with no uncertainty. While this situation cannot be realized in practice, it is useful to study because it allows to easily assess the limiting sensitivity for new counting instruments as a function of the expected background.
 
The second scenario involves an on/off measurement exactly as the one described by \citet{LiMa}, but where there are additional sources of variance, for example systematic uncertainties coming from limited precision in the knowledge of the efficiency of the instrument or from the differences between the background in the off region and the background in the on region. In practical applications uncertainties of this kind are always present, but  are usually neglected. We present in section \ref{sec:lima_sys} a method to account for these additional sources of error when computing the significance and show in section \ref{sec:examples} that these effects can have a non-negligible impact on the significance.

    In the third situation the background estimate $b$ does not come from an off measurement, but from a  model. This model might be for example a polynomial function fitted to off-pulse regions in a time series \citep{GBMCatalog}, a multivariate model trained on historical data \citep{Szecsietal2013,Vasileiou2013} or coming from a simulation. In this case the uncertainty on the prediction of the model becomes the uncertainty in the background estimate $b$, so that $b$ is no longer a Poisson random variable and the formula from \citet{LiMa} cannot be applied. In most cases $b$ is a Gaussian random variable with p.d.f $\mathcal{G}(B, \sigma)$, where $B$ is the true value and $\sigma$ is the error on the estimate. \textbf{Many practitioners apply in this case the formula for the significance $(n - b)/\sqrt{b}$, where $n$ are the counts observed in the on measurement. This assumes that $n$ is a random variable with a Poisson distribution with average $b$, which is only true if $B=b$, i.e., if the uncertainty on the background estimate is negligible ($\sigma \rightarrow 0$). Moreover, it assumes that $b$ is large enough so that the Poisson distribution converges to a Gaussian distribution with variance $\sqrt{b}$. Especially the first condition is seldom true in practice, and can lead to a highly overestimate significance measure}. We derive in section \ref{sec:pois_gaus_sig} a simple analytic expression that can be always used in this case, without the need for further assumptions. To our knowledge, this expression has never been published before.

All the methods presented in this paper are strictly frequentist. For a Bayesian treatment of the on/off measurement problem see for example \citet{Gillessenetal2005}. \textbf{Some hybrid Bayesian-frequentist recipes are also reported in \citet{Cousins2008}.}

\textbf{As a part of this work we publish an open-source Python code \citep[][]{Vianello2018}\footnote{Codebase: \url{https://github.com/giacomov/gv\_significance}} implementing all the methods.}

\section{Model comparison and the Likelihood Ratio Test}
    
In this section we summarize briefly the typical set up of a model comparison test in frequentist statistic, in particular in the context of the detection of a new source or a new effect. We consider a background model $H_{0}$ (\textit{null} hypothesis) and a model $H_{1}$ containing a new source or a new effect (\textit{alternative} hypothesis). The background model comes for example from a side measurement, a simulation, a theory, or some combination of these. We call the parameters of $H_{0}$ and $H_{1}$ respectively $\vec{\theta}_{0}$ and $\vec{\theta}_{1}$ (note that these sets of parameters could also be empty). We further assume that $H_{0}$ and $H_{1}$ are nested, i.e., for each $\vec{\theta_{0}}$ there exists a set $\vec{\theta}_{1}$ so that $H_{0} = H_{1}$ everywhere in the domain of $H_{0}$. It follows that the alternative hypothesis can describe the data at least as well as the null hypothesis for an appropriate choice of its parameters. Therefore, we want to reject $H_{0}$ in favor of $H_{1}$ only if the latter improves on the former \textit{significantly}. To make this idea quantitative, we start by formulating a test statistic (TS), which is a random variable depending on the data and on $H_{0}$ and $H_{1}$. Then we need to know the probability density function for TS under $H_{0}$, which allows to measure the probability $p(\geq TS)$ of obtaining a TS equal \textit{or more extreme} than the TS we measure in our data. This is quoted as the p-value of the test.

In this paper we present simple methods to compute the p-value in counting experiments. These methods are based on the well-known Likelihood Ratio Test (LRT), which has the best performances among the known statistical tools for the problems discussed in this work \citet{Cousins2008}. If $L_{0}(\vec{\theta_{0}})$ and $L_{1}(\vec{\theta_{1}})$ are the likelihood functions for respectively $H_{0}$ and $H_{1}$, the test statistic for LRT is:
$$
TS = 2~\log{ \left( \frac{ \max{\left\{ L_{1}(\vec{\theta_1}) \right\}} }{ \max{\left\{ 
L_{0}(\vec{\theta_0}) \right\}} } \right) }
$$
where $\max{\left\{L(\vec{\theta})\right\}}$ denotes the maximum of the likelihood function. 
We also define the maximum likelihood estimate $\vec{\theta}^{mle}$ of the parameters $\theta$ as 
the parameters which maximize the likelihood function:
$$
\vec{\theta}^{mle} \equiv \argmax_{\vec{\theta}}~L(\vec{\theta}).
$$
It follows that:
\begin{equation}
TS = 2~\log{ \frac{ L_{1}(\vec{\theta_1^{mle}}) }{ L_{0}(\vec{\theta_0^{mle}}) } }
\label{eq:TS_definition}
\end{equation}

An important result from \citet{Wilks} states that, under certain hypothesis, TS in eq.~\ref{eq:TS_definition} is asymptotically distributed as a $\chi^{2}_{d}$, where $d$ is the difference in degrees of freedom between the alternative and the null hypothesis. Under these circumstances the probability $p(\geq TS)$ is simply given by the survival function of the appropriate $\chi^2_d$ distribution.

In physics it is customary to consider, instead of the probability $p(\geq TS)$, the corresponding significance $S$, so that:
$$
\int_{S}^{\infty}~N(x)~dx = p(\geq TS),
$$
where $N$ is the Normal distribution. If Wilks' theorem holds, and if the difference in degrees of freedom between the null and the alternative hypothesis is 1, then $TS$ is a random variable distributed as a $\chi^{2}_{1}$. It follows from the relationship between the Normal and the $\chi^2$ distribution that $\sqrt{TS}$ has a Normal distribution. Therefore, if Wilks' theorem holds, then:
\begin{equation}
S = \sqrt{TS}.
\label{eq:wilks_significance}
\end{equation}
\textbf{In this paper we consider cases where the hypotheses of Wilks' theorem are satisfied. In particular, we consider nested hypotheses and the value of the parameters for which $H_{0} \equiv H_{1}$ are not on the boundaries of their allowed range \citep{Protassov}.}

\section{Detection of a source in a counting experiment}

    Let us consider a counting experiment where we measure $n$ events during an observation (``on'' measurement), and we have an estimate of the expected background $b$ obtained through a side measurement (``off'' measurement) or some other means. In this section we examine the problem of estimating the significance of an effect measured in our experiment. In all the cases examined here, the null hypothesis $H_{0}$ is that we do not have any other effect than the background, characterized by its intensity $B$. The alternative hypothesis $H_{1}$ is that we have a new source characterized by its intensity $M$ on top of the background.

\subsection{Poisson measurement and background with no uncertainty}

    We start from the simplest case, i.e., the ideal situation where we know the expectation $B$ with no uncertainties. This case, however unrealistic, is interesting because the power of the test for this ideal situation can never be surpassed no matter how much we lower the uncertainty on the background.

    In this case we do not need to use a test statistic, as we can compute the probability $p(\geq n)$ of obtaining $n$ counts or more when $B$ are expected by directly summing the Poisson distribution:
$$
p(\geq n) = \sum_{i=n}^{\infty}~\frac{B^{i}~e^{-B}}{i!},
$$
This sum constitute our p-value, and is implemented in most modern statistical libraries. It can be expressed through the $\Gamma$ and the incomplete $\Gamma$ functions:
$$
p(\geq n) = 1 - \frac{\Gamma(n+1, B)}{\Gamma(n+1)},
$$
where $\Gamma(x) = \int_{0}^{\infty}t^{x-1}e^{-t}dt$ and $\Gamma(a, x) = \int_{x}^{\infty}t^{a-1}e^{-t}dt$. 

\begin{figure*}
\centering
\includegraphics[width=0.48\textwidth]{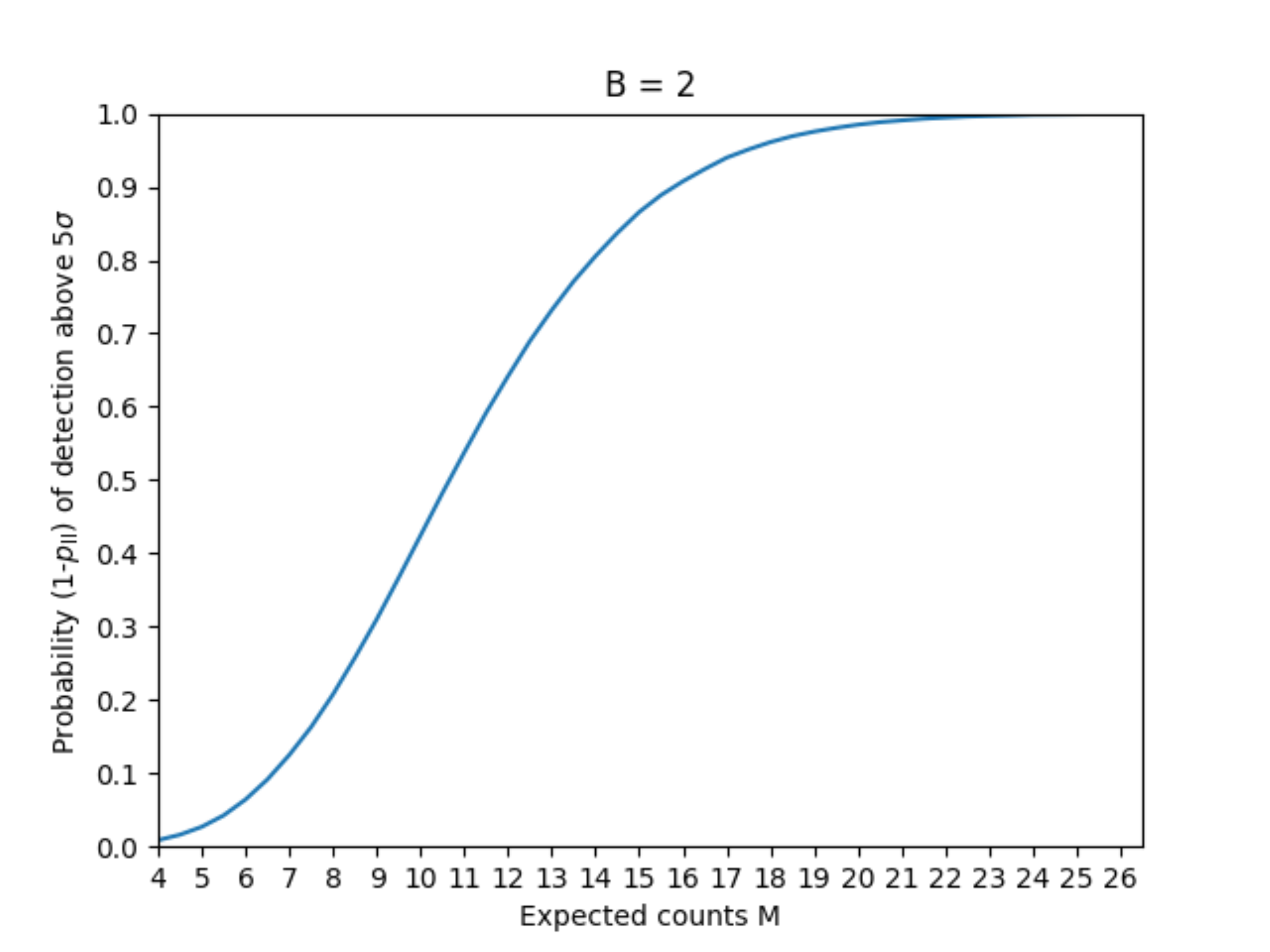}
\includegraphics[width=0.48\textwidth]{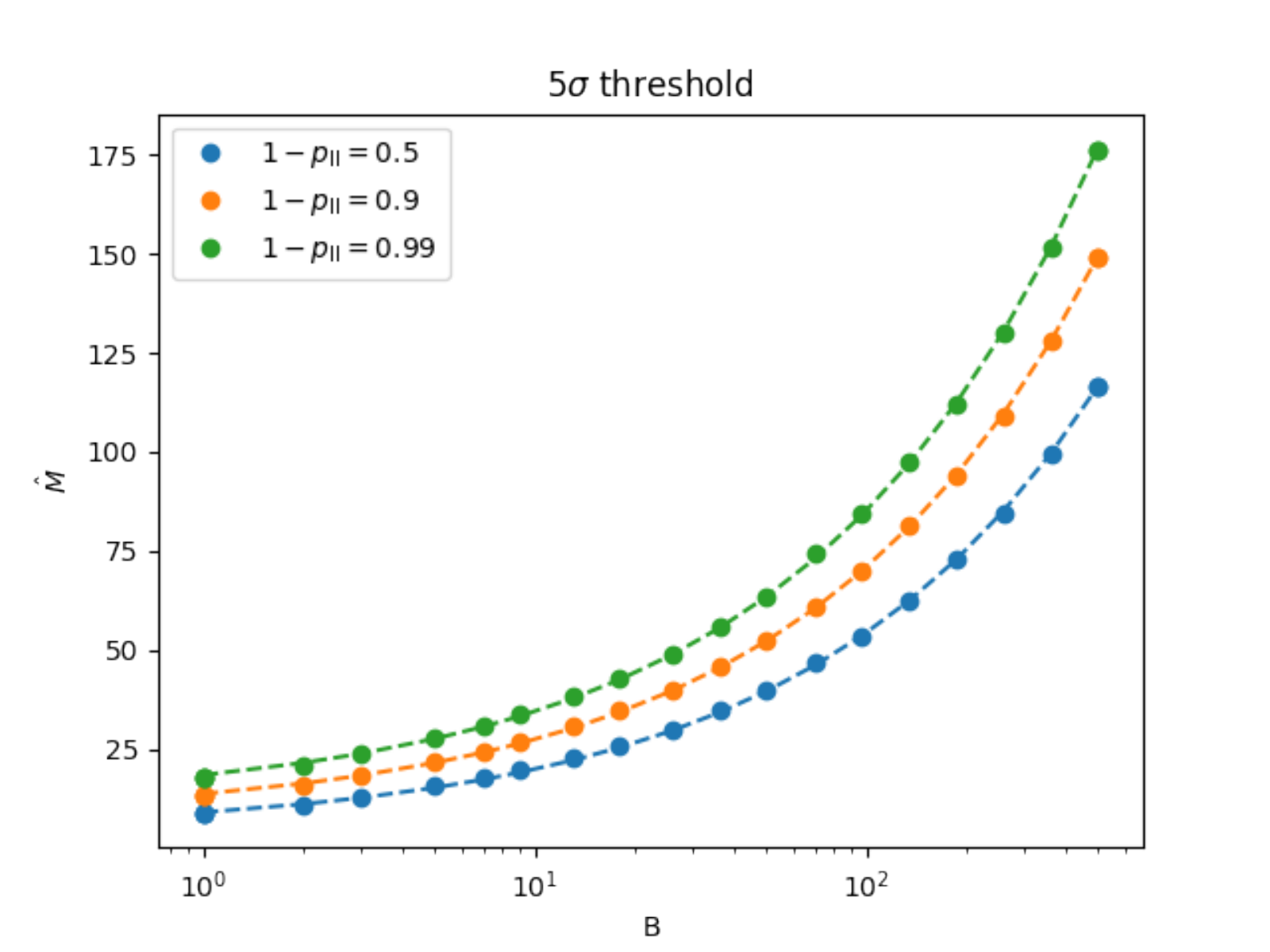}
\caption{\textit{Left panel}: statistical power $1-p_{\rm II}$ as a function of expected counts $M$ for a background of $B=2$. \textit{Right panel}: source counts $\hat{M}$ as a function of $B$ resulting in a efficiency of detection  of 0.5, 0.9 and 0.99 at the $5\sigma$ level.}
\label{fig:stat_power_B2}
\end{figure*}

    Let us now consider a detection threshold of $5\sigma$ (corresponding to a Type I error probability of $p_{\rm I} = 2.86 \times 10^{-7}$), and let us call $p_{\rm II}$ the probability of a Type II error, i.e. of not rejecting the null hypothesis when the alternative hypothesis is true (false negative). Using Monte Carlo simulations we can compute the related statistical power $1-p_{\rm II}$ of the test, i.e. the probability of detecting the new source in one measurement when the alternative hypothesis is true, as a function of $M$ for a fixed $B$. This characterizes the sensitivity of this ideal experiment where the background is known with no uncertainty. It is computed by simulating repeatedly a source with intensity $M$ over a background with intensity $B$ and counting the fraction of realizations for which $p(\geq n) < p_{I}$, and then repeating the procedure for many values of $M$. We show one example for $B=2$ in 
fig.~\ref{fig:stat_power_B2}: a source with $M=10.7$, 15.8 and 20.8 is needed for a probability of detection above $5\sigma$ respectively of 0.5, 0.9 and 0.99. In the right panel of fig.~\ref{fig:stat_power_B2} we show the number of counts $\hat{M}$ as a function of the background $B$ that a source needs to produce in a detector in order to have a probability of 0.5, 0.9 and 0.99 to be detected above $5\sigma$ in one observation. The three curves are approximated with very good accuracy by a function of the form:
\begin{equation}
\hat{M} = a + b ~ \sqrt{B},
\label{eq:m_ideal}
\end{equation}
where the constants are given in table~\ref{tab:coefficients}. This formula can be used in place of Monte Carlo simulations to compute the minimum flux that a source needs to have in order to be detected above $5~\sigma$ with a given efficiency, given an expected background. For example, it can be used to quickly estimate a limit on the sensitivity achievable for a new instruments. The real sensitivity will be worse in any real scenario, when uncertainties on the background are present. \textbf{However, this procedure can be used for example to decide whether it is worthwhile to invest resources into improving background estimation procedure, background rejection, or effective area and to define clear goals for these efforts. We present in section ~\ref{sec:example_ideal} an example of an application.}

\begin{table}
\centering
\begin{tabular}{ c|c|c }
  Detection efficiency & a & b \\
\hline
  $1-p_{\rm II} = 0.5$ &  4.053 & 5.038 \\
  $1-p_{\rm II} = 0.9$ & 7.391 & 6.356 \\
  $1-p_{\rm II} = 0.99$ & 11.090 & 7.415 \\
\end{tabular}
\caption{Coefficients for eq.~\ref{eq:m_ideal}.}
\label{tab:coefficients}
\end{table}

\subsection{Poisson measurement and Poisson background}
\label{sec:pois_pois_sig}

This case has been treated in \citet{LiMa} as well as many other papers \citep[][and references therein]{Cousins2008}. We re-derive it here for completeness, using the formalism we introduced in the previous sections. Let us assume we have performed an off measurement to estimate the background, which returned $b$ counts. Let us call \textit{efficiency} the quantity $e = A~\Delta t$, where $A$ is the effective collecting area and $\Delta t$ is the time during which the instrument was on during the measurement (sometimes called \textit{livetime}). We define $\alpha = e_{on} / e_{off}$, where $e_{on}$ and $e_{off}$ are respectively the efficiency of the experiment and of the side measurement. If the effective collecting area was the same for both measurements, then $\alpha$ is simply the ratio between the exposure times.

The probability of observing $b$ counts during the background measurement is given by the Poisson distribution:
$$
P(b | B) = \frac{B^{b} e^{-B}}{b!},
$$
where $B$ is the ``true'' background, i.e., what we would measure if there was no Poisson noise. Under the alternative hypothesis that there is a new source with intensity $M$, the probability of observing $n$ counts during the source observation is given by:
$$
P(n | M, B) = \frac{(M + \alpha B)^{n} e^{-(M + \alpha B)}}{n!},
$$
where M is the ``true'' source signal. Since the source and background observations are independent, the joint probability of observing at the same time $n$ in the source observation and $b$ in the 
background observation under the alternative hypothesis is simply:
$$
P( n, b | M,B) = P(b | B) \times P(n | M, B).
$$
Taking the logarithm we have:
\begin{align}
& L(n, b | M, B) = n\log{(\alpha B + M)} + b \log{(B)} \nonumber \\
& - (\alpha + 1) B - M,
\label{eq:like1bin}
\end{align}
where we have omitted the term $- \log{(n! b!)}$ because it does not depend on neither B nor M and therefore is inconsequential.
With a little abuse of terminology, we will call this function \textit{likelihood}. Since the alternative and the null hypothesis are nested, the likelihood for the null hypothesis is simply $L(n, b | M=0, B)$:
\begin{align}
& L_0(n, b | B) = n\log{(\alpha B)} + b \log{(B)} \nonumber \\
& - (\alpha + 1) B.
\label{eq:like0bin}
\end{align}
It is easy to find the values for the parameters maximizing $L_{0}$ and $L_{1}$ analytically. The maximum of $L_{0}$ is obtained when:
$$
B^{mle}_0 = \frac{n + b}{\alpha + 1},
$$
while the maximum of $L_{1}$ is obtained when:
$$
B^{mle}_{1} = b,~~M^{mle} = n - \alpha~b.
$$
Substituting these values in eq.~\ref{eq:like1bin} and eq.~\ref{eq:like0bin} and plugging the results in eq.~\ref{eq:TS_definition} we obtain:

\begin{align}
& TS = 2 \left\{ n\log{\left[ \frac{\alpha + 1}{\alpha} 
\left(\frac{n}{o+b}\right)\right]} \right. \nonumber \\
& \left. + b \log{ \left[ (\alpha +1) \frac{b}{n+b} \right] } \right\}.\nonumber
\end{align}

The difference in degrees of freedom between the alternative and the  null hypothesis is 1. Moreover, the hypotheses of Wilks' theorem are satisfied\footnote{Note in particular that since we allow for negative normalizations for the source, the value $M=0$ that reduces the alternative hypothesis to the null hypothesis is not at the boundary of its support \citep{Protassov}.} so that TS is distributed as a $\chi^2$ with 1 d.o.f. and the significance can be written as in eq.~\ref{eq:wilks_significance} as:

\begin{align}
& S = \sqrt{2} \left\{ n\log{\left[ \frac{\alpha + 1}{\alpha} 
\left(\frac{n}{n+b}\right)\right]} \right. \nonumber \\
& \left. + b \log{ \left[ (\alpha +1) \frac{b}{n+b} \right] } \right\}^{1/2},
\label{eq:sign}
\end{align}
which is the expression for the significance found in eq.~17 in \citet{LiMa}.

\textbf{As shown for example in \citet{Cousins2008} this case can also be treated differently by re-writing the likelihood as the product of a Poisson distribution for the sum $B + M$ and the binomial probability that this total is divided as observed, where the Binomial distribution has a parameter $M / (M+B)$. In this case a binomial test can be used \citep{Reid95,James1980,Gehrels86,Clopper94,Zhang1990}. The p-value for the test can be expressed in terms of the \textit{regularized} incomplete $\beta$-function:}
$$
P_{Bi} = \beta_{\rm reg}(o, b, \alpha) \equiv \frac{\beta (o, b, \alpha / (\alpha +1)}{\beta(o, b)},
$$
\textbf{where $\beta(a, b, x)$ is the incomplete $\beta$-function and $\beta (a, b)$ is the (complete) $\beta$-function. Then the significance can be computed as:
\begin{equation} 
Z_{Bi} \equiv \sqrt{2}~Erf^{-1}(1 - 2 P_{Bi}).
\label{eq:z_bi}
\end{equation}
The \textit{regularized} incomplete $\beta$ function $\beta_{\rm reg}$ is implemented in most modern statistical languages\footnote{See for example \url{http://mathworld.wolfram.com/RegularizedBetaFunction.html}}. At the end of section \ref{sec:lima_sys} we will compare the performances of eq.~\ref{eq:sign} and eq.~\ref{eq:z_bi}.}. 

\subsection{Poisson measurement with Poisson background and systematic uncertainty on the background or on the efficiency}
\label{sec:lima_sys}
We assume here a similar situation as in section \ref{sec:pois_pois_sig}. We have a background measurement giving $b$ counts, where the probability density function for $b$ is the Poisson distribution:
$$
P(b | B) = \frac{B^{b} e^{-B}}{b!},
$$
where $B$ is the ``true'' background in the off region. We also have a source observation giving $n$ counts, but we assume this time that the background estimate has a systematic uncertainty. In other words, the true background in the source observation $\hat{B}$ is slightly different from the true background in the background observation $B$. We model this situation by assuming that $\hat{B} = (k + 1) B$, where $k$ is unknown. Of course, if $k=0$ then $\hat{B} = B$, and we must have $k > -1$ because the expected background cannot be negative. In such a situation the distribution for $n$ is:
$$
P(n | M, B) = \frac{\left[M + \alpha (k + 1) B\right]^{n} e^{-\left[M + \alpha (k + 1) B\right]}}{n!}.
$$

The first case we consider corresponds to the practical situation where we have a procedure to perform the background measurement that returns a background counts with limited accuracy. For example, we are considering a measurement where an X-ray telescope has observed the sky and has measured $n$ counts in the source region and $b$ counts in a background region selected around the source region. We may know from domain knowledge that the background is not spatially uniform and in the off region it can be up to 20\% different than the background within the region. Assuming a conservative approach, we can then fix $k=0.2$. The likelihood under the alternative hypothesis in this case is:
\begin{align}
L(n, b | B) = & n \log{\left[B \alpha (k+1) + M\right]} + b \log{B}  \nonumber \\ 
& -B \alpha (k+1) - B - M, \nonumber
\end{align}
where $k$ is fixed a priori. Of course this corresponds to the same situation already treated in section~\ref{sec:pois_pois_sig} where $\alpha \Rightarrow \alpha (k+1)$ or equivalently $B \rightarrow B (k+1)$ (i.e., a systematic uncertainty on the background or on the efficiency have the same effect). The expression for the significance in eq.~\ref{eq:sign} simply becomes:
\begin{align}
& S = \sqrt{2} \left\{ n\log{\left[ \frac{\alpha (k+ 1) + 1}{\alpha (k+1)} 
\left(\frac{n}{n+b}\right)\right]} \right. \nonumber \\
& \left. + b \log{ \left[ \left(\alpha (k+1) +1\right) \left(\frac{b}{n+b} \right) \right] } \right\}^{1/2}.
\label{eq:sign2}
\end{align}
This provides a crude but quick way of penalizing the significance to account for systematic uncertainty on the background or on the efficiency. It also gives an easy way to determine the robustness of a detection. Indeed, we can determine how much changing $k$ affects the significance in eq.~\ref{eq:sign2}. We provide examples of this procedure in section \ref{sec:examples}.

\begin{figure*}[tb!]
\flushright
\includegraphics[width=0.8\textwidth]{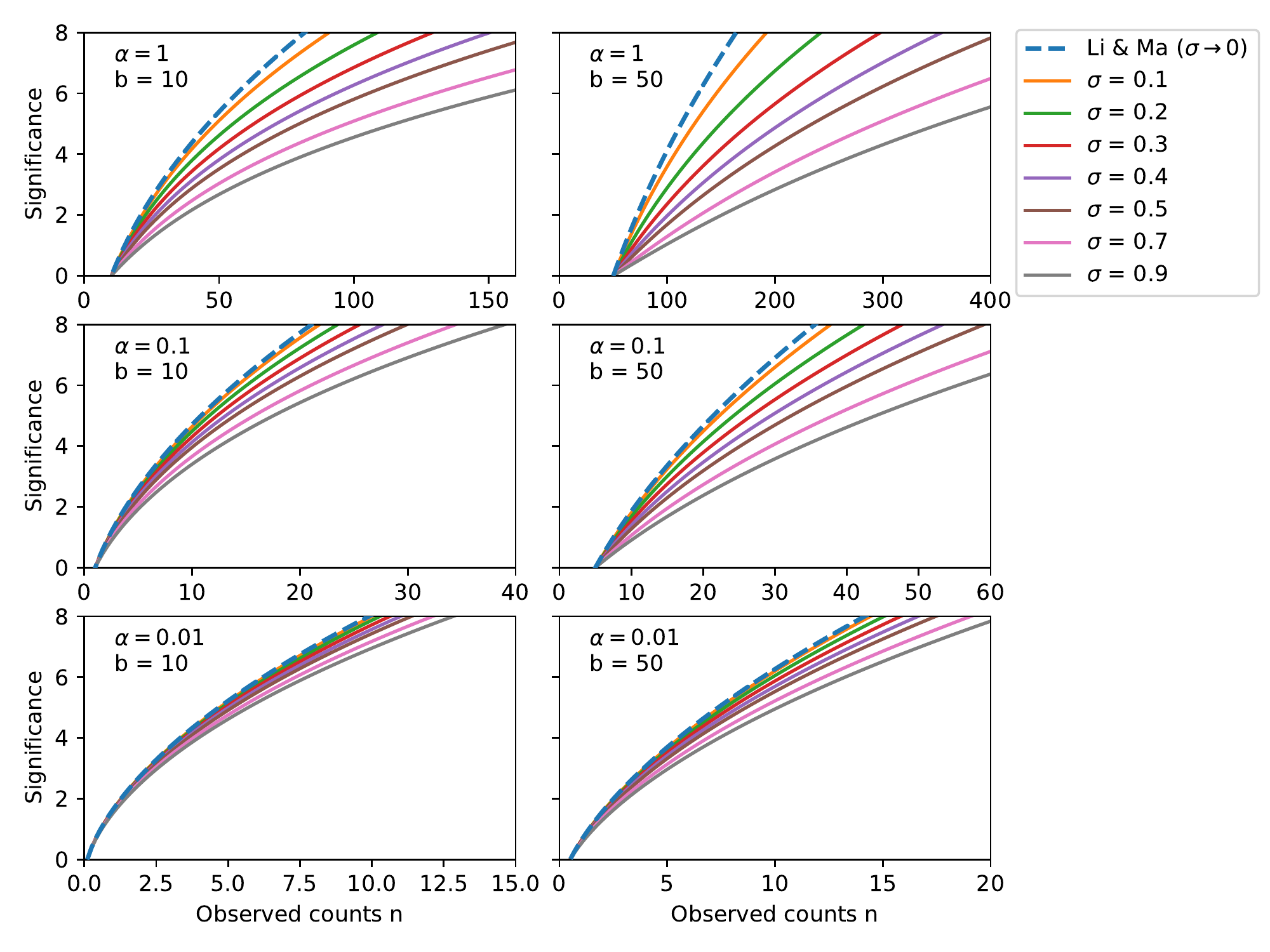}
\caption{Significance as a function of the observed counts $n$ for different $b$ and $\alpha$ values and for different systematic uncertainties $\sigma$.}
\label{fig:sys_errors}
\end{figure*}

A different possibility is to treat $k$ as a random variable. This represents cases when the background or the efficiency estimate (or both) have their own variance. For example, the efficiency is computed through simulations that can only reach a certain accuracy, or the background is measured over a long period of time to accumulate statistic but either the background rate or the detector efficiency are known to vary. In both cases the background or efficiency estimate can be either too low or too high because of the variance of the procedure. We assume that $k$ is a random variable with a Normal probability distribution:
$$
P(k | \sigma) = \frac{1}{\sigma
\sqrt{2\pi}}~\exp{\left[-\frac{k^{2}}{2\sigma^2}\right]},
\label{eq:k_pdf}
$$
where $\sigma$ is the standard deviation and it is a known property of the background estimation procedure. We can then write the joint probability of observing at the same time $n$ in the source observation and $b$ in the 
background observation under the alternative hypothesis as:
\begin{equation}
P( n, b, \sigma | M, B, k) = P(b | B) \times P(n | M, B) \times P(k | \sigma).
\end{equation}
Taking the logarithm and omitting all terms that do not depend on M, B or $k$ we have:
\begin{align}
& L(n, b, \sigma | M, B, k) = n \log{\left[\alpha (k+1) B + M \right]} + b \log{B}\nonumber \\
& - \frac{k^2}{2 \sigma^2} - (k+1)\alpha B - B - M.
\label{eq:like1binsys}
\end{align}
This likelihood was already studied in the past \citep[and references therein]{Spengler2015} in the limit of small systematic uncertainties. Clearly, if $\sigma \rightarrow 0$ so that $k \rightarrow 0$ and $\frac{k^2}{2 \sigma^2} \rightarrow 0$ (i.e., no systematic error) we obtain eq.~\ref{eq:like1bin}.
The maximum of this likelihood is for:
$$
B = b, M = n - \alpha b, k = 0,
$$
and it is:
$$
\max{\left\{L_1\right\}} = b \log{b} - b + n \log{n} - n.
$$

The likelihood for the null hypothesis can be obtained as usual imposing $M=0$:
\begin{align}
& L_0(n, b, \sigma | B, k) = n \log{\left[\alpha (k+1) B \right]} + b \log{B}\nonumber \\
& - \frac{k^2}{2 \sigma^2} - (k+1)\alpha B - B.\nonumber
\end{align}
The solution that maximizes this equation can only be found numerically. However, by equating $\delta L_{0} / \delta B = 0$ we can easily find that for every solution we need to have:
$$
B^{mle} = \frac{b + n}{\alpha (k+1) + 1}.
$$
Therefore, we can substitute $B=B^{mle}$ in eq.~\ref{eq:like0bin} and obtain a maximization problem with only one free parameter ($k$) which is easy to solve numerically. We apply the likelihood ratio test and Wilks' theorem as done in section \ref{sec:pois_pois_sig} and we obtain the following expression for the significance:

\begin{equation}
S = 2^{1/2}~\sqrt{ b \log{b} - b + n \log{n} - n  - \max{\left\{L_{0}\right\}}}.
\label{eq:vianello}
\end{equation}
We provide a code in Python to perform such maximization \footnote{\url{https://github.com/giacomov/gv\_significance}}.
As in the previous case, this equation can also be used to determine how much changing $\sigma$ affects the significance, i.e., how much the result depends on the systematic uncertainties. We give in section \ref{sec:examples} two examples of such a procedure. 

In a real situation where we have a systematic uncertainty of say 10\%, imposing $k=0.1$ in eq.~\ref{eq:sign2} is going to return higher significances than imposing $\sigma=0.1$ in eq.~\ref{eq:vianello}. This is easy to understand, because the p.d.f. for $k$ in eq.~\ref{eq:k_pdf} allows for higher values for $k$ than $k=0.1$. 

\begin{figure*}[tb]
\centering
\includegraphics[width=0.6\textwidth]{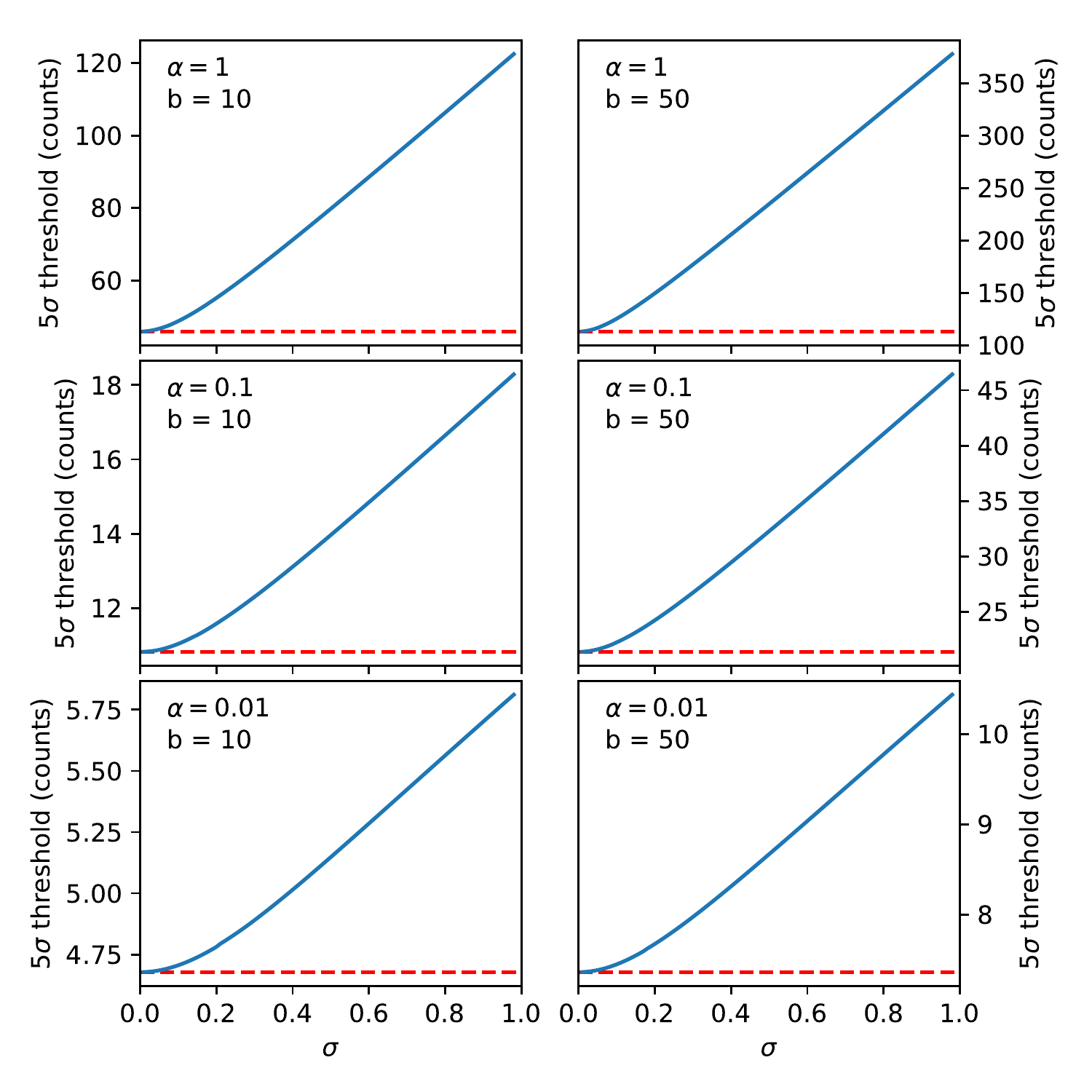}
\caption{Observed counts $n$ needed to get a significance of $5\sigma$ as a function of the systematic uncertainty $\sigma$, for different values of $b$ and $\alpha$. The red dashed line is the threshold obtained with the formula from \citet{LiMa}, i.e., the case with no systematic errors.}
\label{fig:5sigma_threshold}
\end{figure*}

In Fig.~\ref{fig:sys_errors} we show some examples of the significance as a function of $n$ obtained by applying this formula for different values of $\sigma$, $b$ and $\alpha$ (colored lines). We also report for comparison the significance for the case with no systematic uncertainty (eq.~\ref{eq:sign}, blue dashed line). As expected, increasing $\sigma$ increases the number of observed counts $n$ required to reach a given significance. In Fig.~\ref{fig:5sigma_threshold} we show the counts corresponding to the $5\sigma$ threshold for different $b$ and $\alpha$ as a function of the systematic error. We note that after an initial shallow increase, the threshold increases linearly with $\sigma$. The linear part corresponds to the regime where the systematic uncertainty on the background estimation dominates over the statistical error.

\begin{figure*}[tb!]
\flushright
\includegraphics[width=0.8\textwidth]{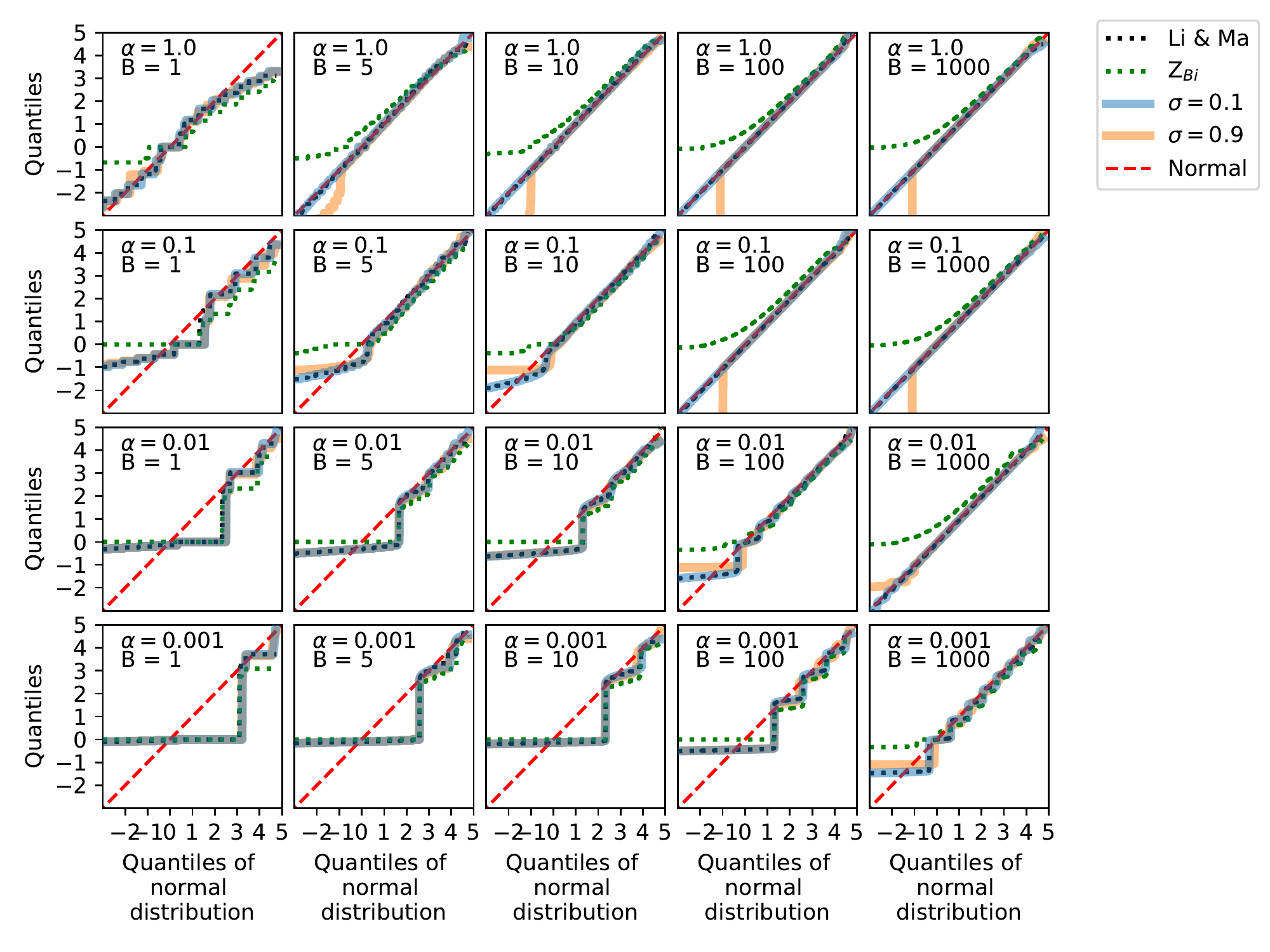}
\caption{Quantile-quantile plots obtained through Monte Carlo simulations and using eq.~\ref{eq:vianello} for different values of $\alpha$ and $B$. See text for details.}
\label{fig:qqplot}
\end{figure*}

In order to study the range of applicability of eq.~\ref{eq:vianello}, as well as the performance of eq.~\ref{eq:sign} and eq.~\ref{eq:z_bi} presented in section \ref{sec:pois_pois_sig}, we performed Monte Carlo simulations. We have performed in particular 1 million simulations of the null hypothesis (M=0) for different values of $\alpha$, $B$ and $\sigma$. For this study we assign arbitrarily a sign of $-1$ to the significance when $n < b$, and a sign of $+1$ otherwise. Under this conditions the probability density function of the significance should be a Normal distribution. We show in Fig.~\ref{fig:qqplot} the quantile-quantile plot for different cases of $\alpha$ and $B$ and for $\sigma=0.1$ (blue datasets) and $\sigma=0.9$ (yellow datasets), as well as for the case with no systematic error, representing eq.~\ref{eq:sign} with black dots and eq.~\ref{eq:z_bi} with green dots. In this kind of plot, if the distribution of $S$ is indeed a Normal distribution, it should align with the diagonal (red dashed line). We can immediately see that this is not the case for $B=1$ (left panels) independently of the formula used, although some formulae are closer to the diagonal. Also, we can see that negative significances (which we have assigned when $n < b$) are in general not well behaved. The reason is easy to understand: $n$ and $b$ are counts, thus they have a lower bound at 0 while they have no upper bound. Therefore, downward fluctuations are constrained while upward fluctuations are not. For all these reasons, eq.~\ref{eq:vianello}, eq.~\ref{eq:sign} and $Z_{Bi}$ should not be used for quantifying the significance of under fluctuations unless $\alpha B$ is large and $\sigma$ is small. This is generally not a problem, given that we are interested in the discovery of new effects above the background, and not below. We also note that there appears to be steps in the quantiles when $\alpha B \lesssim 1$. These are due to the discrete nature of the Poisson distributions for $n$ and $b$, and do not constitute a problem. \textbf{When considering the case with no systematic errors ($\sigma = 0$), eq.~\ref{eq:z_bi} (green dotted line) appears to be overly conservative for small $B$ with respect to eq.~\ref{eq:sign} (black dotted line). The latter is too conservative only for the extreme case in the upper left corner (but still less than $Z_{Bi}$) and should therefore be preferred in the regimes considered here. When considering the case with systematic errors ($\sigma > 0$), for $B \ge 5$ eq.~\ref{eq:vianello} works well for all positive significances.}

\subsection{Poisson measurement and Gaussian background}
\label{sec:pois_gaus_sig}
Here we consider the case where the background estimate $b$ is a random variable with a Normal probability density distribution, instead of a Poisson distribution as in the previous cases. This happens often in practice when the background estimate $b$ does not come from a side measurement but from a background model or a procedure which returns $b$ and its standard error $\sigma$. For example, let us consider the problem where we want to estimate the significance of a signal in a time series. A common methodology adopted for the study of Gamma-Ray Bursts \citep{GBMCatalog} starts by selecting a time window around the signal (pulse window), and two off-pulse time windows respectively before and after the time window of interest. A polynomial function is then fitted to the off-pulse windows and used to estimate the expected background $b$ within the pulse window. In this case the uncertainty on $b$ is not described by Poisson statistic. Instead, $b$ is a random variable with a Gaussian p.d.f. $\mathcal{G}(B, \sigma)$, assuming that the polynomial fit is well-conditioned and that there are enough data in the off-pulse window to constrain the parameters of the polynomial. The standard deviation $\sigma$ of $\mathcal{G}$ can be estimated by propagating the errors on the parameters of the polynomial (optionally adding a systematic contribution if necessary). \textbf{This case was treated previously in \citet{Cousins2008}. Here we add on that work by explicitly deriving an analytic formula that can be easily applied in this case.}

Let us then assume that we have observed $n$ counts during the observation of a source of interest, and that we have a method for estimating the background which returns an expected value $b$ with standard deviation $\sigma$. The p.d.f. for $b$ is then a Gaussian distribution:
\begin{equation}
P(b | B,\sigma) = \frac{1}{\sigma
\sqrt{2\pi}}~\exp{\left[-\frac{(b-B)^{2}}{2\sigma^2}\right]},
\label{eq:gauss_b}
\end{equation}
where $B$ is the true background value.
Under the alternative hypothesis, the distribution for the $n$ counts is the Poisson distribution:
$$
P(n|M,B) = \frac{ (M + B)^{n}~e^{-(M+B)} }{n!}.
$$

We can proceed as in the previous sections by writing the joint probability for $n,b$ under the alternative hypothesis as:
$$
P(n,b|M,B,\sigma) = P(n|M,B)~P(b | B,\sigma),
$$
and taking the logarithm:
\begin{align}
L(n, b, \sigma | M, B) = & -\frac{(b-B)^2}{2 \sigma^2} + n \log{(B+M)} \nonumber \\
& - B - M.
\label{eq:gausslike}
\end{align}
The likelihood for the null hypothesis is obtained by imposing $M=0$:
\begin{equation}
L_{0(n, b | B, \sigma)} = -\frac{(b-B)^2}{2 \sigma^2} + n \log{(B)} - B.
\label{eq:gausslike0}
\end{equation}
The maximum of this expression is obtained for:
$$
B^{mle}_0 = \frac{1}{2} (b - \sigma^2 + \sqrt{b^2 - 2 b \sigma^2 + 4 n \sigma^2 + \sigma^4}),
$$
\textbf{where we have chosen the positive solution for $B^{mle}_{0}$ since the true value of the background cannot be negative.}
We note here that the p.d.f. for $b$ in eq.~\ref{eq:gauss_b} allows for $b<0$, which might seem unnatural. However, what really cannot be negative is not $b$ but $B$, i.e., the \textit{true} value of the background. The expression for our best estimate $B^{mle}_0$ is consistent with this expectation, since it is never negative even if $b<0$ given that $\sigma > 0$ and $n \ge 0$. 
The maximum for eq.~\ref{eq:gausslike} is obtained for:
$$
B^{mle}_1 = b, M^{mle} = n - b.
$$

\begin{figure*}[tb!]
\flushright
\includegraphics[width=0.8\textwidth]{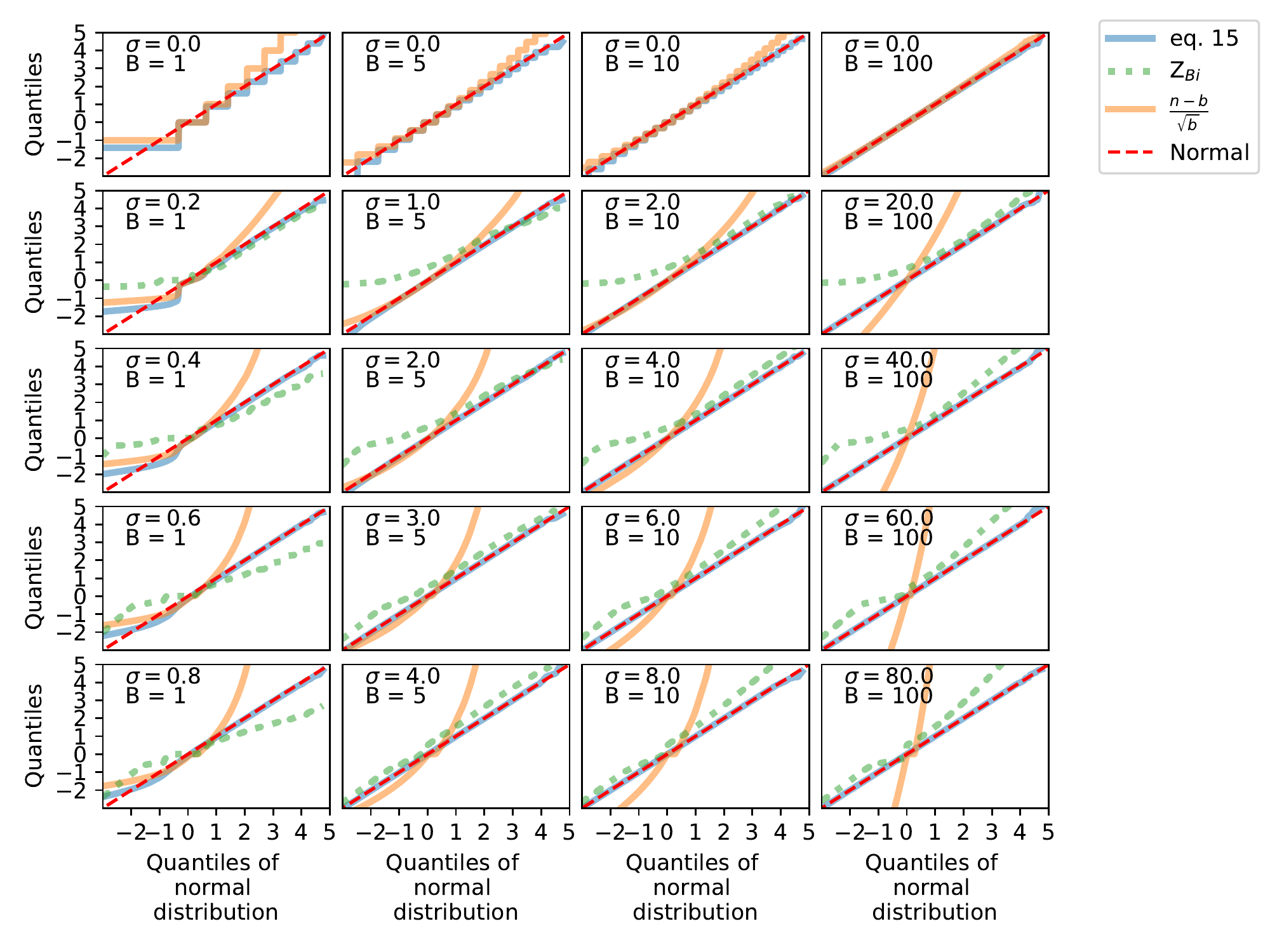}
\caption{Quantile-quantile plots obtained with Monte Carlo simulation and eq.~\ref{eq:poigau_significance}. See text for details.}
\label{fig:qqgau}
\end{figure*}

Substituting in eq.~\ref{eq:gausslike0} and eq.~\ref{eq:gausslike} and using eq.~\ref{eq:TS_definition}, the test statistic becomes:
$$
TS = 2 \left[ n \log{\left(\frac{n}{B^{mle}_{0}}\right)} + \frac{(b - B^{mle}_{0})^2}{2 
\sigma^2} + B^{mle}_{0} - n  \right].
$$
Once gain, the difference in degrees of freedom between the null and the alternative hypothesis is 1, so we can use eq.~\ref{eq:wilks_significance}
and the significance is:
\begin{equation}
S = \sqrt{2} \left[ n \log{\left(\frac{n}{B^{mle}_{0}}\right)} + \frac{(b - B^{mle}_{0})^2}{2 
\sigma^2} + B^{mle}_{0} - n  \right]^{\frac{1}{2}}.
\label{eq:poigau_significance}
\end{equation}

In order to explore the range of applicability of eq.~\ref{eq:poigau_significance} as a function of $B$ and $\sigma$ we have performed Monte Carlo simulations. In particular, we simulated 1 million realizations $(n, b)$ for each of a set of different values of $B$ and $\sigma$. In these simulations we have assigned a negative sign to $S$ when $n<b$ and a positive sign otherwise. We show the results in the quantile-quantile plots in Fig.~\ref{fig:qqgau}, where we demonstrate that the quantiles of the distribution of $S$ in eq.~\ref{eq:poigau_significance} (blue line) are very close to the quantiles of the Normal distribution (diagonal, red dashed line), as expected. We note that there are ``steps'' in the upper right panels, due to the fact that the Poisson distribution for $n$ is discrete and therefore the significance jumps from one level to the next when $n$ increases by one. There are also some features for negative significances in some of the panels. These are due to the fact that counts are bounded to be positive or zero, therefore they cannot oscillate too far in the downward direction especially when $B$ is small. This means that the significance returned by eq.~\ref{eq:poigau_significance} should not be taken seriously when $n < b$ and $b$ is small. This does not constitute a problem normally as we are interested in detecting sources above the background (and a source cannot have negative flux by definition).
\textbf{In the same figure we also report the results for a formula which is often used in the literature, namely $S=\frac{n-b}{\sqrt{b}}$. This expression neglects the uncertainty on the background estimate (i.e., it assumes $\sigma=0$ and hence $B=b$) and it assumes that $b$ is large enough that the Poisson distribution converges to a Gaussian distribution. This is the case in the upper right panel in Fig.~\ref{fig:qqgau}. However, in all other cases these assumptions are violated and that significance is largely overestimated}. Therefore, we argue that practitioners should instead use eq.~\ref{eq:poigau_significance}, which does not require further assumptions and works better.

\textbf{\citet{Cousins2008} provide also an alternative recipe for this case based on an approximate equivalence with eq.~\ref{eq:z_bi} where $\alpha = \sigma^2 / b$ and $b \rightarrow b / \alpha$. This approximation is represented by the green dotted line in fig.~\ref{fig:qqgau}. It works well in some cases (see for example the left panels in the second row) but it overestimate the significance in some regimes and underestimate it in some others.}

\section{Examples}
\label{sec:examples}
In this section we examine some simple examples of the application of the formulae provided in this paper.

\subsection{A faint short Gamma-Ray Burst}

\begin{figure*}[tb]
\centering
\includegraphics[width=0.47\textwidth]{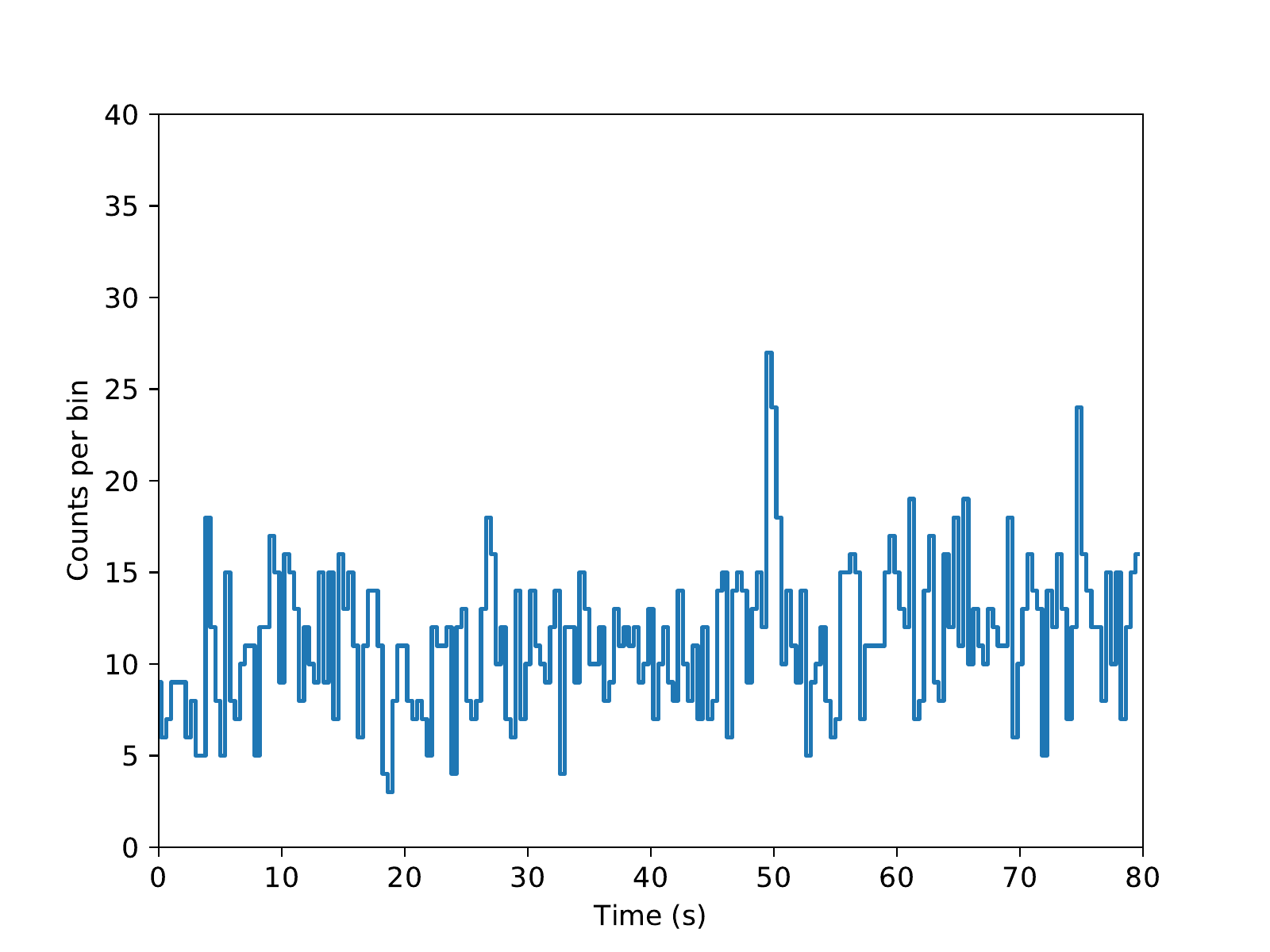}
\includegraphics[width=0.47\textwidth]{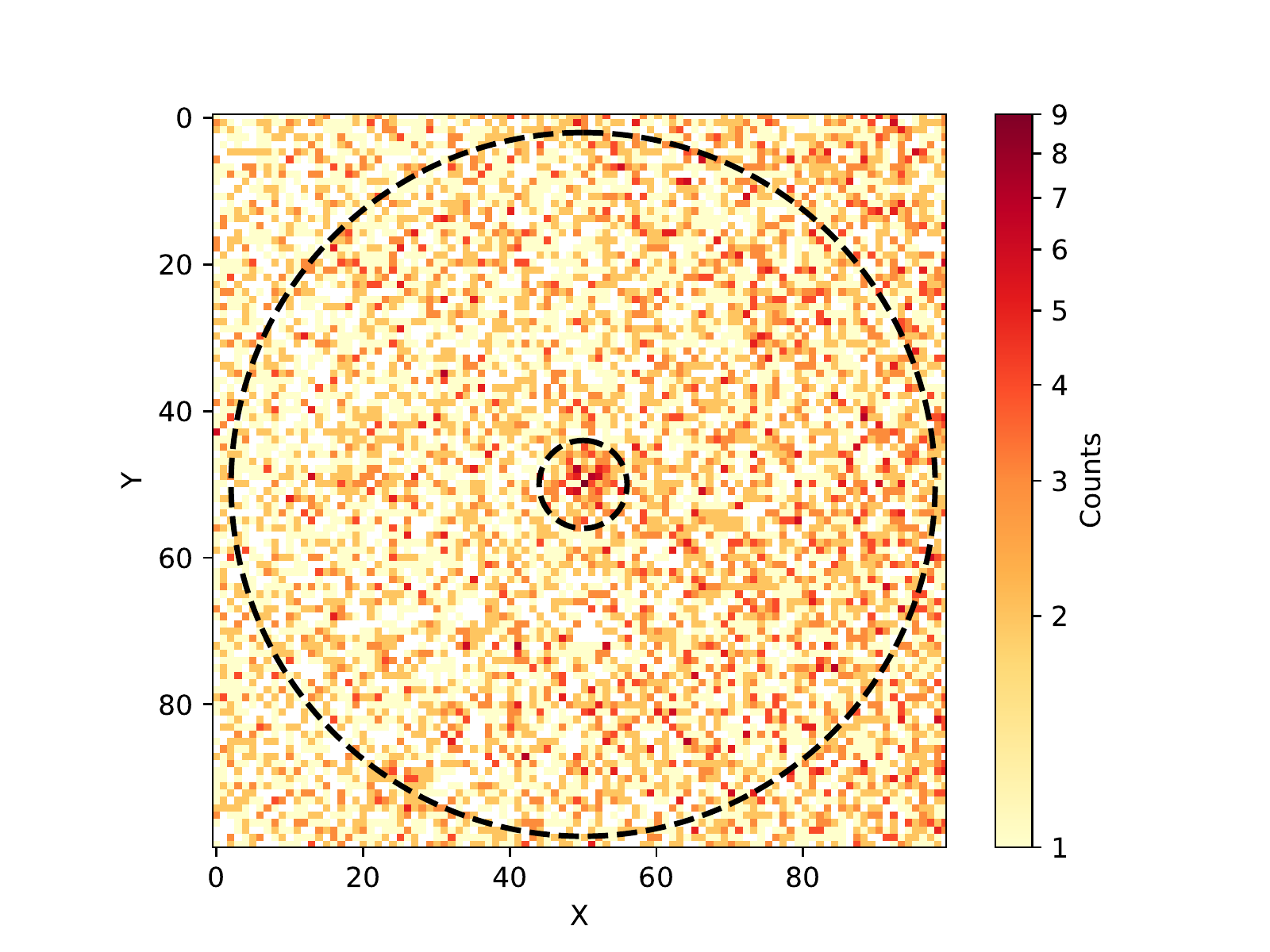}
\caption{\textit{Left panel}: simulation of a faint short Gamma-ray Burst and a background component as seen by a counting detector (no imaging). \textit{Right panel}: simulation of a source (center) on a slightly non-uniform background. The source region is marked by the inner black circle, while the background region is the annulus between the inner and the outer black circle.}
\label{fig:grb_sim}
\end{figure*}

We consider here a typical case for the detection of a source in a time series. In the left panel of Fig.~\ref{fig:grb_sim} we show a simulation of a light curve of a short Gamma-Ray Burst superimposed to a slowly varying background. This is a typical situation for counting detectors such as the Fermi Gamma-ray Burst Monitor \citep{theGBM}), with no imaging capabilities. We can see a candidate short signal around 50 s and we want to determine its significance. The easiest way is to select an off-pulse window (for example between 0 and 40 s) and an on-pulse window (from 49.4 to 50.6) and use the formula from Li\&Ma (eq.~\ref{eq:sign}). Using the notation of section~\ref{sec:pois_pois_sig} we have $n=69$, $b=1046$ and $\alpha = 0.03$, and we obtain a significance of $S = 5.7$. However, using eq.~\ref{eq:sign2} we can determine that an increase in the background of just 10\% ($k = 0.1$) changes the the significance to $S = 5$, barely above the $5\sigma$ threshold, and $k=0.2$ gives $S=4.5$. This means that an in-depth study of the systematic uncertainties on the background is in order, because our result is sensitive to these uncertainties. For example, we need to show that we can keep them $\lesssim 10$\% in order to claim a detection at the $>5\sigma$ level. Now let us consider the case where we have a background estimation procedure, for example like the one in \citet{Szecsietal2013}, and that we know from validation studies that it gives a background estimate with a typical error of 10\%. We can then use eq.~\ref{eq:vianello} with $\sigma = 0.1$ to estimate the significance, and obtain $S=4.9$. We also might note that there seems to be a small increasing trend in the background light curve. A widely-used alternative in a case like this is to use a second off-pulse window (for example between 55 s and 80 s), fit a line to the two off-pulse windows and then interpolate the line in the on-pulse window to obtain an estimate of the background which accounts for the trend \citep[see for example]{GBMCatalog}. We therefore consider the model $a~t + b$ and we fit it to the off-pulse windows by maximizing a Poisson log-likelihood \citep{Cash}, obtaining $a = (4.6 \pm 1.1) \times 10^{-2}$ and $b = 9.5 \pm 0.5$. By propagating the errors we obtain an estimate for the background counts in the on-pulse window of $b=35.4 \pm 0.9$ counts. This measurement is with good approximation a random variable with a Gaussian distribution, therefore we cannot apply eq.~\ref{eq:sign} but we need instead eq.~\ref{eq:poigau_significance} with $b=35.4$ and $\sigma=0.9$, which yields $S=4.9$, significantly smaller than the significance obtained with eq.~\ref{eq:sign} and similar to the significance obtained assuming a systematic error of $10$\%. Unfortunately, many practitioners use in these cases the approximated formula $S = (n-b) / \sqrt{b}$ which neglects the uncertainty on the background estimate and assumes $b$ large enough (see section \ref{sec:pois_gaus_sig}). This would yield in this case the anti-conservative estimate $S=5.8$. Our eq.~\ref{eq:poigau_significance} does not make any approximation and accounts for the uncertainty on the background estimate, and therefore should be preferred in all cases where $n$ is a Poisson random variable and $b$ has uncertainty, no matter whether $b$ is small or large.

\subsection{A source in imaging data}

Here we consider a typical case for X--ray or $\gamma$-ray astronomy, where we have an image taken from an imaging detector and we are looking for point sources. In the right panel of Fig.~\ref{fig:grb_sim} we show a simulated image of such an observation where we can see a slight excess in the center, and we want to estimate its significance. We can start by using the formula from \citet{LiMa} (eq.~\ref{eq:sign}). \textbf{We consider a circle centered on the source with a radius chosen as to contain a large fraction of the Point Spread Function of the instrument (for example 99\% of the encircled energy)}. We consider the counts contained in such a circle (inner black circle in fig.~\ref{fig:grb_sim}) as the on source measurement, and the counts contained in the annulus between the inner and the outer circle as off measurement.  We have $n=296$, $b=12301$, and the ratio between the area of the two circles is $\alpha=0.0159$. Using eq.~\ref{eq:sign} we obtain $S=6.6$. We note, however, that the intensity of the background increases slightly moving from the left to the right. Therefore, the background estimate likely has an additional systematic uncertainty. Eq.~\ref{eq:sign2} shows that an increase in the background level of just 10\% ($k=0.1$) lower the significance to $S=5.2$ and an increase of 15\% ($k=0.15$) gives $S=4.5$, thus we need to show that we can keep the systematic uncertainty on the background $\lesssim 15$\% to claim a detection at the $>5\sigma$ level. \textbf{Let us now consider the case where we have a procedure to estimate the background, for example like the one in \citet{Vasileiou2013}, which counts events in a properly chosen background region over a long period of time. Let us suppose it gives a measurement of $b=123010$ with an $\alpha = 0.00159$ and a fractional systematic uncertainty of $\sigma = 0.15$ mainly due to subtle time variations of the background. Using eq.~\ref{eq:vianello} we find $S=3$, a large difference with respect to the initial estimate.} Using the same equation we find that we need to reduce the systematic error on the background estimate to at least $\sigma = 0.06$ in order to claim a significance $>5\sigma$. Alternatively, we can model the background as a plane inclined from left to right, i.e., $c_i = a + b~X_i$ where $c_i$ are the counts in the i-th spatial bin and $X_i$ is its x coordinate. We can fit this function to the data by maximizing a Poisson likelihood as in the previous section, and obtain an estimate of the background counts expected in the inner circle (the source region). In particular we obtain $b = 192.95 \pm 9.7$. The measurement $b$ is with good approximation a random variable with a Gaussian distribution, therefore we cannot use eq.~\ref{eq:sign} anymore, and we need instead to use eq.~\ref{eq:poigau_significance}, obtaining $S=5.6$, still considerably lower than the significance obtained directly with the \citet{LiMa} formula in eq.~\ref{eq:sign}.

\subsection{A sensitivity study}
\label{sec:example_ideal}
\textbf{In this example we estimate whether a current or future observatory will be capable of observing sources from a given class, starting from an estimate of the typical flux. Let us consider a very simple case, and imagine that we are studying a future non-imaging instrument sensitive in the range 10 keV - 1 MeV and we want to know whether it would be able or not to detect a signal from a source like GRB 170817A - the first Gamma-Ray Burst associated with a Gravitational Waves events \citep{Abbott2017}. The flux of the GRB was measured by Fermi/GBM to be $(2.8 \pm 0.2) \times 10^{-7}$ erg cm$^{-2}$ and its duration was $\sim 2$ s in the energy range of our instrument \citep{Goldstein2017}. Let us suppose we have already a design of the instrument and an estimate of the expected background of 5 events per second, i.e., $B=10$ events over the duration of the signal. Using eq.~\ref{eq:m_ideal} we find that we need 19.98, 27.49 and 34.54 photons from the source in order to detect it above $5\sigma$ respectively in 50\%, 90\% and 99\% of the cases, in the ideal case of no background uncertainty. Since we have a design of the instrument we also typically have an estimate of its effective area, and by convolving the spectrum measured by Fermi/GBM with the effective area we obtain an estimate of the expected signal over the duration of the event. If we obtain an expected photon fluence of say $F=100$ we can immediately conclude that we will have no problems in detecting such a source. On the other hand, if we obtain $F=20$ photons we can conclude that our instrument is not sensitive enough to detect the source, no matter the precision of our background estimation procedure. If possible, we need to go back to the design phase and increase the effective area by at least 1.5-2 times. If however we obtain $F=30$ photons it means that we are very close to our ideal sensitivity limit, and we would expect to detect a source such as GRB 170817 with a little more than 90\% efficiency if we had no uncertainty on the background. Hence, we should invest effort in studying the impact of background uncertainties on our efficiency and demonstrate that we can keep them small enough. In particular, using Monte Carlo simulations and the formulae provided in this work we can determine the maximum tolerable Gaussian uncertainty for estimation methods providing Gaussian errors, or the maximum tolerable factor $\alpha$ and systematic uncertainty $k$ for on/off methods, as a function of the detection efficiency. We already know, however, that such efficiency will never be larger than the boundary fixed by the ideal case ($\sim 90$\%), no matter the accuracy of our background estimation procedure.}

\section{Conclusions}
In this paper we have provided techniques to account for and to assess the importance of systematic uncertainties when measuring the significance of a source or effect. We have also provided for the first time a simple formula to compute the significance in the case where the observed counts are a Poisson random variable but the background is a Gaussian random variable (eq.~\ref{eq:poigau_significance}), and not a Poisson random variable as assumed by the classic formula in \citet{LiMa} (eq.~\ref{eq:sign}). This typically happens when the background estimate comes from a model which has been fit to archival or ancillary data. We have also provided a simple formula to compute the number of counts that a source should produce in a detector to have a probability of being detected of 50\%, 90\% or 99\% above $5\sigma$ when the background is known perfectly (eq.~\ref{eq:m_ideal}). This constitute the maximum sensitivity that a counting instrument can achieve, \textbf{and can be used for simple studies on the sensitivity of instruments to specific classes of sources}. In section~\ref{sec:examples} we have shown three examples which illustrate how to use the different formulae. These examples also demonstrates that ignoring additional uncertainties on the background estimate can yield an overestimated significance in fairly common circumstances.

\section{Acknowledgements}
The author thanks Professor Robert Cousins (UCLA), Professor James T. Linnemann (MSU) and the anonymous referee for the useful discussion and for providing comments that helped improving considerably the paper.


\begin{thebibliography}{dummy}
\bibitem[Abbott et al.(2017)]{Abbott2017} Abbott, B.~P., Abbott, R., Abbott, T.~D., et al.\ 2017, \apjl, 848, L13 
\bibitem[Cash(1979)]{Cash} Cash, W.\ 1979, \apj, 228, 939
\bibitem[Clopper (1934)]{Clopper94} Clopper, C.~J., Pearson, E.~S.\ 1934, Biometrika, 26, 404
\bibitem[Cousins et al.(2008)]{Cousins2008} Cousins, R.~D., Linnemann, J.~T., \& Tucker, J.\ 2008, Nuclear Instruments and Methods in Physics Research A, 595, 480
\bibitem[Gehrels (1986)]{Gehrels86} Gehrels, N.\ 1986, \apj, 303, 336 
\bibitem[Gillessen \& Harney(2005)]{Gillessenetal2005} Gillessen, S., \& Harney, H.~L.\ 2005, \aap, 430, 355 
\bibitem[Goldstein et al.(2017)]{Goldstein2017} Goldstein, A., Veres, P., Burns, E., et al.\ 2017, \apjl, 848, L14 
\bibitem[James \& Roos(1980)]{James1980} James, F., \& Roos, M.\ 1980, Nuclear Physics B, 172, 475 
\bibitem[Li \& Ma(1983)]{LiMa} Li, T.-P., \& Ma, Y.-Q.\ 1983, \apj, 272, 317 
\bibitem[Meegan et al.(2009)]{theGBM} Meegan, C., Lichti, G., Bhat, P.~N., et al.\ 2009, \apj, 702, 791-804 
\bibitem[Narayana Bhat et al.(2016)]{GBMCatalog} Narayana Bhat, P., Meegan, C.~A., von Kienlin, A., et al.\ 2016, \apjs, 223, 28 
\bibitem[Protassov et al.(2002)]{Protassov} Protassov, R., van Dyk, D.~A., Connors, A., Kashyap, V.~L., \& Siemiginowska, A.\ 2002, \apj, 571, 545 
\bibitem[Reid (1995)]{Reid95} Reid, N.\ 1995, Statistical Science, 10, 138
\bibitem[Spengler(2015)]{Spengler2015} Spengler, G.\ 2015, Astroparticle Physics, 67, 70 
\bibitem[Sz{\'e}csi et al.(2013)]{Szecsietal2013} Sz{\'e}csi, D., Bagoly, Z., K{\'o}bori, J., Horv{\'a}th, I., \& Bal{\'a}zs, L.~G.\ 2013, \aap, 557, A8 
\bibitem[Vasileiou(2013)]{Vasileiou2013} Vasileiou, V.\ 2013, Astroparticle Physics, 48, 61 
\bibitem[Vianello(2018)]{Vianello2018} Vianello, G.\ 2018, gv\_significance, v.1.0.0, Zenodo, \href{https://doi.org/10.5281/zenodo.1157308}{doi:10.5281/zenodo.1157308}
\bibitem[Wilks (1938)]{Wilks} Wilks, S.S.\ 1938, Ann. Math. Statist., 9, 60
\bibitem[Zhang \& Ramsden(1990)]{Zhang1990} Zhang, S.~N., \& Ramsden, D.\ 1990, Experimental Astronomy, 1, 145 
\end{thebibliography}
\end{document}